\documentclass[twocolumn,preprintnumbers,amsmath,amssymb]{revtex4}

\usepackage{graphicx}
\usepackage{amssymb}
\usepackage{psfrag}
\usepackage{url}
\usepackage{hyperref}
\newcommand\beq{\begin{equation}}
\newcommand\eeq{\end{equation}}
\newcommand\bea{\begin{eqnarray}}
\newcommand\eea {\end{eqnarray}}

\begin{document}

\title{The OPERA neutrino velocity result and the synchronisation of clocks}

\author{Carlo R.~Contaldi}
\affiliation{Theoretical Physics, Blackett Laboratory, Imperial
  College, SW7 2BZ, London, United Kingdom}
\date{\today}

\begin{abstract}
  The CERN-OPERA experiment \cite{opera} claims to have measured a
  one-way speed of neutrinos that is apparently faster than the speed
  of light $c$. One-way speed measurements such as these inevitably
  require a convention for the synchronisation of clocks in
  non-inertial frames since the Earth is rotating. We argue that the
  effect of the synchronisation convention is not properly taken into
  account in the analysis of \cite{opera} and may well invalidate
  their interpretation of superluminal neutrino velocity.
\end{abstract}

\maketitle

The CERN-OPERA experiment attempts to measure the velocity of
neutrinos generated at CERN and received at the Laboratorio Nazionale
del Gran Sasso (LNGS). The method requires accurate knowledge of the
distance baseline traveled by the neutrino along with detailed
budgeting of any delays affecting the electronic transmission
of timing signals at both ends of the baseline. The overall time of
flight is established by time stamping local reference clocks at both
sites using a single, {\sl common view} GPS clock
signal received separately at both sites. The time difference
between time stamped clock signals at either end is calibrated using a
second atomic clock transported between the two GPS receivers.

The stated result, after allowing for all budgeted time delays, is
that the Time Of Flight (TOF) of the neutrinos is shorter by $60 \,
\pm 6.9\,(\mbox{stat.})\, \pm 7.4\,(\mbox{sys.})$ ns than that expected if
they traveled at the speed of light $c$. If true, such a claim would
have profound consequences for our understanding of fundamental
physics.

Various authors have already commented on more or less exotic Lorentz
violating theories for the neutrino that may account for the result
\cite{Cacciapaglia:2011ax,AmelinoCamelia:2011dx,Autiero:2011hh,Ciborowski:2011ka,Alexandre:2011kr,Klinkhamer:2011mf,Giudice:2011mm,Dvali:2011mn,Gubser:2011mp}. Here
we address the question of whether the stated measurement of the
neutrino velocity is self-consistent.

One-way speed of light measurements in non-inertial frames of
reference have always suffered from an interpretation problem due to a
requirement for synchronising two or more clocks \cite{Zhang}. The
problem does not exist in two-way speed of light measurements in which
the (light) signal is reflected back to the origin of the apparatus
where its arrival is measured with the same clock that measured its
departure.

In this {\sl letter} we will focus on the problem of synchronisation
of the time signals used to calculate the TOF of the
neutrino beam by the OPERA experiment. We will argue that there is
sufficient ambiguity in the synchronisation and as such, the result
may not be interpreted as done in \cite{opera}.

The synchronisation of clocks is a fundamental problem in accelerating
frames as only inertial observers are equivalent in general
relativity. The OPERA experiment {\sl attempts} to get around
this problem by time-stamping their time chains using the clock signal of
a single GPS satellite. GPS satellite signals have relativistic
corrections applied to their tick rate and transmission
frequencies. The aim of the corrections is to make the time signal as
close as possible to a Universal Time Coordinate (UTC) equivalent to
the time coordinate of an inertial frame of reference encompassing the
Earth and all GPS satellites. The corrections applied counter the
Sagnac effect due to the Earth's rotation, the time
dilation due to the position of the satellites in the Earth's
gravitational potential and some second order corrections (see 
\cite{gps} for an extended review). The largest of these corrections are of ${\cal O}(10^2)$
ns. The definition of UTC used by GPS satellites is determined by the United States
Naval Observatory (USNO) and the clock of any single GPS satellite is managed to be, on average,
within about 100 ns of its arbitrary definition of the inertial frame
UTC \footnote{\url{http://www.usno.navy.mil/USNO/time/gps}}.  

Due to the remaining discrepancy with UTC and both receiver and
atmospheric transmission effects it is still not possible to use the
GPS signal received at the two ends of the baseline as synchronous UTC
markers with the required ns precision. This appears to be consistent
with the $\sim 100$ ns accuracy estimate quoted by \cite{Ellis} when
considering at what level OPERA could detect Lorentz violations.  In
an effort to go beyond this accuracy threshold, the OPERA
experiment employed a travelling {\sl Time-Transfer Device} (TTD) to
calibrate the difference in time signals at each receiver. We assume
this device to be a transportable atomic clock of sufficient accuracy
\footnote{At the time of writing some of the references in
  \cite{opera} concerning these measurements refer to publications
  that are not public or lead to recursive http links.}. The TTD
constitutes a classic moving clock synchronisation conundrum in
relativity.

The problem lies in the fact that the TTD was moved between baseline
ends in an accelerating frame due to the Earth's rotation. At first
order in $1/c^2$ this introduces {\sl three} relativistic time
distortions which will induce a discrepancy between the clock at the
CERN end of the baseline where the TTD was initially synchronised and
the the time shown by the TTD when it is compared to the clock at
LNGS. The first effect is the time dilation due to moving the TTD
through a non-uniform gravitational potential. This effect is
exacerbated by the non-inertial frame as it makes it path dependent as
opposed to dependent only on the potential difference between the
starting and end points. The second is a Doppler type effect due to
the velocity $v$ of the TTD with respect to the Earth's rotating frame
of reference. The third is the Sagnac effect due to the rotation of
the Earth as the TTD travels to its destination. All three effects
must be integrated along the path ${\cal C}$ taken by the TTD en-route
to LNGS. The time discrepancy accumulated by the TTD over the path
${\cal C}$ can be stated as \cite{gps}
\begin{eqnarray}\label{eq:dt}
  \Delta t &\sim & -\int_{\cal C}
  \frac{(V-V_{\rm CERN})}{c^2}\,d\tau + \nonumber\\ 
  &&\frac{1}{2}\int_{\cal C}\frac{v^2}{c^2}\,d\tau+2\frac{\omega_E}{c^2}\int_{\cal C} \, dA_z\,,
\end{eqnarray}
where $\tau$ is the proper time along the path, $dA_z$ is the
infinitesimal area swept out by the vector connecting the (polar)
$z$-axis of the reference frame to the TTD position along the path and
projected onto the equatorial plane, and $\omega_E = 7.2921151467\times
10^{-5}$ s$^{-1}$ is the rotation rate of the Earth. Of the remaining
terms, $V$ is the gravitational potential at a given radius $r$ and
latitude $\theta$ and $V_{\rm CERN}$ is a reference value of the
potential at the starting point. The rotating Earth is not a perfect sphere and
a quadrupolar model for the effective gravitational potential of the
Earth including a centripetal contribution is
given by
\begin{equation}
  V = -\frac{GM_E}{r}\left[1-J_2\left(\frac{r_E}{r}\right)^2P_2(\cos \theta)\right]+\frac{1}{2}(\omega_E\,r)^2\,,
\end{equation}
where $G$ is Newton's constant, $M_E$ is the mass of the Earth with
$GM_E=3.986004418\times 10^{14}$ m$^3$ s$^{-2}$, $r_E=6.378137\times 10^{6}$ m,
$P_2$ is Legendre's polynomial of second order, and $J_2=1.0826300\times
10^{-3}$ is the Earth's quadrupole moment. The radius $r$ can be
obtained from the geoid model for the equipotential surface and is
described by a polynomial expansion in $x=\sin \theta$ as
\begin{equation}
 r \sim 6356742.025 + 21353.642\,x^2+39.832\,x^4\ \ \mbox{m}\,,
\end{equation}
to fourth order in $x$. At CERN's approximate latitude we have $V_{\rm
  CERN}/c^2 = -6.951546823\times 10^{-10}$ and the difference between
potentials at CERN and LNGS is $\Delta V/c^2 = 7.82 \times 10^{-14}$.

Let us now make some simple assumptions about the path ${\cal C}$
taken by the TTD on its way from CERN to LNGS; that is that the TTD
moved at a constant speed between the two ends of the baseline and the
journey took 12 hours. We will assume the journey followed a radius
from the origin given by the quadrupolar model for the Earth's shape
between the two latitudes. 

Of the three effects included in (\ref{eq:dt}) only the time dilation
effect acts in such a way that the TTD runs slower than the reference
clock at CERN and could therefore explain the anomalously low TOF
observed by OPERA if the time stamp at OPERA were calibrated using the
TTD. The time dilation estimated along the simple trajectory turns out
to be $\Delta t \sim 2$ ns. The Sagnac effect acts in the opposite
direction with a value of $\Delta t\sim 4$ ns whilst the Doppler
effect is negligible at these speeds. At first glance therefore it
appears that the overall $\Delta t$ by which the TTD is shifted from
the CERN clock is of ${\cal O}(1)$ ns in the direction {\sl opposite}
to what is required to explain the anomaly. However, whilst the Sagnac
effect only accumulates as the TTD is moving, the time dilation effect
accumulates even when the TTD is stationary and at a value of the
potential which differs from that at CERN. As an example of how this
can further affect the synchronisation procedure, let us assume that
the TTD was stationary at the LNGS site for 4 days while the apparatus
for clock comparison was set up. Using the value of $\Delta V/c^2$
quoted above this would result in a total shift of $\Delta t \sim 30$
ns.

We therefore argue that the true time-link discrepancy for the two
baseline ends should be a (negative) time dilation effect that can
easily be of ${\cal O}(10)$ ns plus a (positive) Sagnac effect of
${\cal O}(1)$ ns \footnote{We have adopted the same sign convention as
  \cite{opera} such that a negative correction will alleviate the
  measured discrepancy.}. The quoted time-link discrepancy of 2.3 ns
in \cite{opera} may well be irrelevant since it is obtained by
comparing the asynchronous LNGS reference clock with the TTD which, as we have
argued, is no longer synchronised with the CERN clock and is shifted
by amounts comparable to those estimated here by the time a comparison
is made at LNGS. Such a comparison has no quantitative meaning if the
details and timings of the journey that the TTD undertook from CERN up to the
clock comparison event at LNGS are not taken into account.

It is worth noting that the time dilation effect is sensitive to many
aspects of the journey taken by the TTD. Even disregarding the
acceleration phases of such a journey which would move the TTD between
distinct non-inertial frames, the effect depends strongly on the
velocity of the TTD during the trip and its altitude above the Earth's
surface - was it transported by car or aeroplane? Did it sit for hours
in a baggage terminal? Did the TTD stop for lunch at a scenic
spot in the Alps? etc. It may seem counter-intuitive that such details of
a non-relativistic journey could lead to appreciable effects but the
time differences the result hinges on are extremely small.

Additional effects that exacerbate the path dependence of the time
dilation effect is the deformation of the gravitational potential
beyond the simple quadrupole model along the route taken by the
TTD. These have not been taken into account here but may be relevant
particularly around the mountainous regions along the TTD route.

More importantly, we have only considered the path taken by the TTD
along a surface trajectory. The path taken by the neutrinos is some 3
kms below the surface at its midpoint along the trajectory connecting
CERN and LNGS. At this level of accuracy the surface time measured by
all clocks involved will differ from the proper time along the true
trajectory and this further complicates the interpretation of the
OPERA results.

In conclusion, a number of effects have not been taken into account
when considering the synchronisation convention adopted by the OPERA
experiment. The simple estimates given in this work show that the time
difference expected after transporting the TTD between baseline ends
invalidate its use as a synchronising master clock at the required
level of precision. Although \cite{opera} does not state explicitly
that these effects have not been taken into account, it is hard to
imagine how a calculation as elaborate as that required to compute
$\Delta t$ accurately would not have been referred to in their
discussion. At the very least, some discussion of the time line
between synchronisation at CERN and comparison at LNGS should be
included in their analysis. It appears that OPERA has fallen foul of
the same stumbling block of past one-way speed of light measurements
in non-inertial frames. The resulting measurement that the neutrino
velocity differs from $c$ is not only unsurprising but should be {\sl
  expected} in their setup.

{\sl Acknowledgements} We thank Toby Wiseman for useful
discussions. The author acknowledges the hospitality of CTCRM
Lympstone where some of this work was initiated. We also acknowledge
Sgt Jamie Sanderson RM for first having pointed out the CERN-OPERA
result to the author and for engaging discussions about the mysteries of
the Universe we live in.

\bibliography{letter}
\end{document}